\documentclass[11pt]{article}
\usepackage{amssymb}
\usepackage{amsmath}
\usepackage{graphicx}
\usepackage{dcolumn}
\usepackage{bm}

\begin{document}

\title{Condon domains -\\
these non-magnetic diamagnetic domains}
\author{\textnormal{Valerij S. Egorov} \\
\\
{\small {\textit{Russian Research Centre "Kurchatov Institute"}}}\\
{\small {\textit{Kurchatov Sq. 1, Moscow 123182, Russia }}}\\
{\small {\textit{e-mail: egorov@isssph.kiae.ru}}}\\
{\small {\textnormal{Received 22 April 2003, accepted 17 August 2004}}}}
\date{}
\maketitle

\begin{abstract}
The paper, not pretending for a complete and detailed review, is intended
mainly for a wide community of physicists, not only specialists in this
particular subject. The author gives a physical picture of the periodic
emergence of instabilities and well-known diamagnetic domains (Condon
domains) in metals resulting from the strong de Haas-van Alphen effect. The
most significant experiments on observation and study of the domain state in
metals are described. In particular, the recent achievements in this area
using muon spin rotation $\mu $, as well as the amazing phenomenon of
``supersoftness'' observed in the magnetostriction experiments, are
presented. Novel, not previously discussed features of the phenomenon
related to the metal compressibility are enlightened. 

The paper is based on lectures given by the author at \'{E}cole 
Doctorale de Physique (graduate school of physics) in Grenoble 
(France) during a stay with Institut National Polytechnique de 
Grenoble.\newline
\newline
\textbf{PACS}: 75.45.+j; 71.70.Di; 75.60.d
\end{abstract}

\bigskip

Actually, the title contains no contradiction. The term ``non-magnetic'' is
to emphasize the absence of a connection between the phenomena to be
discussed and the magnetic moments of atoms causing such well known magnetic
phenomena as para- ferro- and antiferromagnetism, magnetic domains, etc. We
will deal with simple metals with a zero atomic magnetic moment where
unrestricted motion of conduction or free electrons is the only source of
magnetism. The motion of free electrons in a magnetic field is known to be
circular, due to the Lorentz force; the projection of electron trajectory
onto a plane normal to the magnetic field, forms a closed \textit{Larmor
orbit}, and this orbital, \textit{diamagnetic }motion (since the sign of
Larmor orbit magnetic moment is always negative) causes a peculiar
magnetization of a metal with formation of \textit{diamagnetic domains.} Its
peculiarity consists in the fact that this magnetization, known as the de
Haas -- van Alphen (dHvA) effect, occurs in all metals but only at very low
temperature, very high uniformity of a fairly strong magnetic field, and
very high quality of a metallic \textit{single crystal.} Moreover, for
observation of diamagnetic domains, first predicted by J.H. Condon [1], all
these conditions become more severe.

At first glance, the above contains some hidden contradiction. Indeed, the
Larmor orbit is diamagnetic at any temperature so that the lower is the
magnetic field, the higher is its negative magnetic moment, while the field
uniformity seems to be irrelevant at all. On the other hand, the magnetic
field only \textit{bends} the trajectory of an electron, not changing its
energy. From this \textit{classical} point of view, if the electron energy
does not change in the magnetic field, it is senseless to magnetize only
increasing the energy in vain. It is just the case, and the contradiction
with negative magnetic moments of Larmor orbits has a very simple
explanation. The case is that simultaneously with the high diamagnetic
moment caused by electron rotation in the bulk of metal, some part of
electrons, which are closer to the metal surface than the Larmor diameter,
can no longer form a closed orbit, running into the surface. These
electrons, bouncing from the surface, move on average in the opposite
direction yielding a positive magnetic moment, and create a \textit{%
paramagnetic }effect, exactly compensating the diamagnetism of all internal
electrons. We will try to demonstrate this result in a simplest way.

Let us consider a plane containing electrons with the surface density $N$
rotating in a magnetic field along the circular orbits of the radius $R=v$ /$%
\omega $. Here $v$is the constant velocity of electrons and $\omega $ is the
cyclotron frequency. We cut now a square of the size $a\gg R$. The total diamagnetic moment of all electrons in the
square is%
\begin{equation*}
M{\_}=1/cNa^{2}J_{o}S_{o}.
\end{equation*}%
Here $c$ is the light velocity, $J_{o}=\omega e$/2$\pi $ is the current of
one electron on the Larmor orbit, $S_{o}=\pi R^{2}$ is the orbit are\textbf{.%
} 

\begin{figure}
   \centering
   \includegraphics[width=105mm,height=34mm]{fig1.gif}
   \caption{Cutting of Larmor orbits at the sample edge. Here, $R$ is the Larmor orbit radius, y is the distance from orbit centers to the sample edge, $L$ is the length of charge transfer for a given orbit along the edge.}
    \label{Fig1}
\end{figure}

Thus we have%
\begin{equation*}
M{\_}=1/2cNa^{2}\omega eR^{2}.
\end{equation*}%
The compensating paramagnetic moment is the result of electron's moving
along edge cutting orbits. All orbits, that have a distance $y$ between
their center and the cutting line ${\left| y\right|< R}$ (see Fig. 1), are cut and hence the number of cutting circles $n$ is%
\begin{equation*}
n=4a2RN=8NaR.
\end{equation*}%
The average value $L_{av}$ of the shift $L(y)$ we find (see Fig. 2) as%
\begin{equation*}
L_{av}\mathbf{\mathit{\ }}=\frac{1}{2R}\int_{R}^{-R}L(y)dy\mathbf{\mathit{.}}
\end{equation*}%

\begin{figure}
   \centering
   \includegraphics[width=55mm,height=37mm]{fig2.gif}
   \caption{More detailed picture of a cutting orbit.}
    \label{Fig2}
\end{figure}

Replacing $y=R$cos$\alpha $\textit{, dy}%
\textbf{\ = }$R$sin $\alpha d\alpha $\textbf{\textit{\ }}and $L(y)$ = 2$R$
sin $\alpha $\textbf{\textit{,}} we obtain%
\begin{equation*}
L_{av}\mathbf{\mathit{\ }}=\mathbf{\mathit{\ }}\frac{1}{2R}\int_{0}^{\pi }{%
2R^{2}\sin ^{2}\alpha d\alpha }=R\int_{0}^{\pi }{\sin ^{2}\alpha d}\alpha 
\mathbf{\mathit{.}}
\end{equation*}%
On the interval (0,$\pi )$%
\begin{equation*}
\int_{O}^{\pi }{\sin ^{2}\alpha d\alpha \equiv \int_{0}^{\pi }{\cos
^{2}\alpha d\alpha }},\text{ so that }L_{av}\mathbf{\mathit{\ }}=\mathbf{%
\mathit{\ }}\frac{\pi R}{2}.
\end{equation*}

\begin{figure}
   \centering
   \includegraphics[width=53mm,height=38mm]{fig3.gif}
   \caption{Two symmetrical orbits with the same $L$, situated \textquotedblleft inside\textquotedblright\ and \textquotedblleft outside\textquotedblright\ of a sample.}
    \label{Fig3}
\end{figure}

We find now the average velocity of electrons along the edge $v_{a}$ by
combining the two cutting orbits of length $\alpha $ and \textbf{\textit{\ }}%
$2\pi -\alpha $ (see Fig. 3). They both have the same value of shift $L$ and
whole time of moving on them is exactly the period $T=2\pi \mathbf{/}\omega 
\mathbf{\mathit{.}}$\textbf{\textit{\ }}So, the velocity $v\mathbf{=}2L/T$
and the average velocity is%
\begin{equation*}
v_{av}=2L_{av}/T=\omega R/2.
\end{equation*}%
Thus, an average electron turns around the whole edge for the time $t^{\ast
}=4a\mathbf{/}v_{av}\mathbf{=}8a/\omega R$ and the current of one electron
is $e/t^{\ast }=e\omega R/8a$. Now we have to remember that the number of
cutting circles is $n$ and every circle before cutting contains only one
electron. Of course, the electron after cutting can find itself either
inside or outside cutting line. Since the number of electrons is very high,
the probability for electron to put itself inside the cutting line, i.e., in
our square, is exactly one half. So, we have the number of skipping
electrons exactly $n/2$, the whole paramagnetic current is

\begin{equation*}
I_{+}=\frac{ne}{2t^{\ast }}=\frac{1}{2}Ne\omega R^{2},
\end{equation*}%
\begin{equation*}
M_{+}=1/2cNa^{2}e\omega R\mathbf{^{2},}
\end{equation*}%
and%
\begin{equation*}
M_{+}=M{\_}.
\end{equation*}

Therefore, non-magnetic materials should remain completely non-magne-tic. It
was, however, found long ago that a number of them, particularly bismuth,
graphite, and some other, demonstrate a noticeable diamagnetism. It means
that in those metals magnetic field can by some means increase the electron
energy. But how can it be done?

L.D. Landau was the first who considered this problem from the quantum, or
wave, mechanical point of view. From this point, a free moving particle can
be associated with some fixed wavelength \textit{$\lambda $} (de Broglie
wavelength) inversely proportional to the particle momentum. It is clear
that for a free particle motion, \textit{$\lambda $}, as well as the
particle momentum and energy, can by, in general, arbitrary. However, if the
motion is confined by a so-called potential box, then, roughly speaking, an
integer number of wavelengths must be kept within a box. This means that 
\textit{$\lambda $} can no longer be an arbitrary, continuously varying
parameter. Respectively, the particle energy can also change only by fixed
portions, \textit{quanta.} Of course, a piece of metal also represents a
potential box for conduction electrons moving in it but its dimensions are,
as a rule, so large that none of the electrons can cross it for its
\textquotedblleft free life\textquotedblright , or relaxation time $\tau $,
which is the period between collisions with defects or impurities,
inevitably present even in a very pure metal. For this reason, we can for
sure neglect the size quantization. At the same time, the size of a Larmor
orbit, inversely proportional to the magnetic field, is, as a rule,
essentially less than dimensions of a real metallic sample, so that the
probability of impurity or defect scattering at this orbit in a good sample
is fairly low, especially in high magnetic fields. In other words, in this
case the relaxation time $\tau $ is much more than the period of Larmor
orbit 2$\pi $/$\omega $, i.e. $\omega \tau \gg 1$, and the electron motion
at this orbit can be considered as a closed, \textit{finite }one.

This approach brought L.D. Landau in 1930 [2] to the idea of equidistant,
the so-called \textit{Landau levels}. In the standard electron energy vs
momentum dependence

\begin{equation*}
E = \frac{1}{2m}p^2 = \frac{1}{2m}(p_x^2 + p_y^2 + p_z^2 ),
\end{equation*}

\noindent the energy is a continuous function of any projection of the
momentum $\eth $. In the magnetic field $H$, it can be presented in the form

\begin{equation*}
E=\frac{1}{2m}(p_{\bot }^{2}+p_{\shortparallel }^{2}),
\end{equation*}

\noindent where $p_{\bot }$ and $p_{\shortparallel }$ are, respectively,
normal and parallel projections of the momentum \textbf{\textit{$\eth$}} on
the magnetic field direction. Instead of it, Landau obtained a principally
new result:

\begin{equation*}
E=\left( n+\frac{1}{2}\right) \hbar \omega +\frac{1}{2m}p_{\shortparallel
}^{2}.
\end{equation*}

Here $n$ is an integer acquiring the values 0, 1, 2, \ldots up to a some
maximal one, $\hbar $ is the Planck's constant divided by $2\pi ,\omega
=eH/mc$ is the cyclotron frequency, that is the frequency of electron
rotation in a magnetic field, $m$ is the electron mass, and $c$ is the light
velocity. In this case, the energy of electrons moving along closed orbits
in the plane perpendicular to the magnetic field directions, can no longer
change continuously. It changes by fixed portions, \textit{quanta}, which
magnitude $\hbar \omega $ is proportional to the magnetic field strength. It
is essential that the minimal electron energy begins not from zero but $%
\hbar \omega /2$. At the same time, electron motion along the magnetic field
remains unchanged.

Landau showed that the total energy of such quantized electron gas exceeds
its classical value by a correction proportional to $H^{2}$, resulting in a
negative magnetization, linear in magnetic field, and thus explaining the
diamagnetism of free electrons. Besides this result, Landau found that for
the magnetic field values large enough compared with the temperature, i.e.~if%
\begin{equation}
\hbar \omega \gg kT
\end{equation}%
($k$ is the Boltzmann constant), the field dependence of magnetic moments
becomes essentially non-linear. The magnetic moment vs field dependence
acquires a \textquotedblleft fast periodicity\textquotedblright , or
magnetization oscillations. In essence, it was a prediction of a new
phenomenon. Unfortunately, at that time no ideas exist of a great variety of
the Fermi surface shapes and sizes in metals, and the free electron model
Landau based on, yielded extremely high requirements to the magnitude and
uniformity of magnetic field, practically unachievable at that time, and he
expressed a doubt concerning feasibility of experimental observation of this
effect. Nevertheless, field oscillations of the magnetic moment with the
period inversely proportional to the magnetic field, were soon discovered in
bismuth by de Haas and van Alphen [3] and got the name of de Haas - van
Alphen (dHvA) effect.

Later, the dHvA effect was observed in other metals as well but the
oscillations were seen only in high quality single crystals and at very low
temperatures. The oscillation amplitude dropped fast even at small
temperature increase. The period of oscillations appeared to vary widely in
different metals with the difference reaching several orders of magnitude.
In some metals several periods were observed almost simultaneously (see Fig.~4), 
and the value of periods depended, as a rule, on the crystal orientation
related to the magnetic field direction. It is not surprised that for a
rather long time the magnetization oscillations were not directly associated
with the Landau's prediction.

\begin{figure}
   \centering
   \includegraphics[width=89mm,height=38mm]{fig4.gif}
   \caption{Example of dHvA oscillations with two frequencies in inverse magnetic field. The small period is about 10 times smaller than big one.}
    \label{Fig4}
\end{figure}

Such a versatility of experimental results managed to be understood only
later, on the base of the LAKP (I.M. Lifshits, M.Ya. Azbel, M.I. Kaganov,
V.G. Peschanskii) theory describing versatility of the Fermi surface shapes
and sizes. In 1952 L. Onsager demonstrated first [4] that the constant
period of magnetization oscillations as a function of inverse magnetic
field, is inversely proportional to the area $A$ of extremal cross-section
of the Fermi surface by a plane perpendicular to the magnetic field
direction (see Fig.~5). The bigger is the area $A,$ the \textquotedblleft
faster\textquotedblright\ the magnetization oscillates. For instance, the
dHvA from a Fermi surface like the dumbbell with two cross-sections is shown
on Fig. 4. The inverse period -- \textit{magnetic frequency }-- is given by
the Onsager relation

\begin{equation*}
F=\frac{c\hbar A}{2\pi e}.
\end{equation*}

\noindent Eventually, in 1955, I.M. Lifshits and A.M. Kosevich [5] developed
the theory of metal magnetization (LK-theory). The authors advanced much
further than Landau, obtaining a result adequate for any metal with
arbitrary Fermi surface at arbitrary temperature, which, for the free
electron model, naturally coincided with that of Landau. However, only due
to the LK theory it became clear why the Landau diamagnetism might me
anomalously large in several and why it is practically independent of
temperature.

\begin{figure}
   \centering
   \includegraphics[width=94mm,height=71mm]{fig5.gif}
   \caption{Different shapes of Fermi surfaces (schematically): spherical, long ellipsoid, dumbbell. The extremal belt is shown. The wider is a belt, the stronge is an oscillating contribution in energy and, consequently, dHvA amplitude.}
    \label{Fig5}
\end{figure}

Progress in theory served as a kind of impact for an unprecedented growth in
the number of experimental investigations of metals in magnetic field at low
temperature [6]. The measurements of magnetization oscillations have become
one of the basic methods to study Fermi surface. During one decade, an
enormous number of works was done and for all or almost all metals, at least
those for which high quality single crystals could be grown, Fermi surfaces
were ``decoded''. And just here we are eventually approaching the main
subject of this paper. It appears that importance of the dHvA effect is not
restricted to its ``benefits'' in the Fermi surface decoding. The
oscillating field dependence of the energy of metals in magnetic field is
the base of some remarkable low-temperature phenomena being of independent
interest. The formation of diamagnetic domains is definitely one of them.

So, let the formation of Landau levels in external uniform magnetic field $H$
cause in a metallic sample some oscillating addition $\tilde{\varepsilon}$
to the energy and, respectively, oscillating magnetization (the dHvA
effect). This means that the magnetic field inside the sample, or \textit{%
magnetic induction }$B$, differs slightly from the external field $H$\textit{%
.} It is this difference%
\begin{equation*}
B-H=4\pi M
\end{equation*}

\noindent which does represent the oscillating magnetic moment. Thus,
besides $\tilde{\varepsilon}$\textbf{, }we should also bear in mind the
energy of excess magnetic field $B-H$ in the sample\textbf{\textit{.}}
Taking for simplicity our sample in the form of a long cylinder parallel to
the magnetic field, we can write the total energy change per unit volume as
the sum

\begin{equation}
\tilde{\varepsilon}+(B-H)^{2}/8\pi .  \label{eq1}
\end{equation}

Since $\tilde{\varepsilon}$ is determined by the magnetic field $B$ acting
on electrons, and oscillates in this field, it is evident that $B$ will
change relative to $H$ always towards the nearest minimum of $\tilde{%
\varepsilon}$\textbf{. }The exact value of $B$ is obtained from the
obligatory condition for this sum to acquire its minimal possible value,
which requires vanishing its $B$-derivative. It means that

\begin{equation*}
\frac{\partial \tilde{\varepsilon}}{\partial B}+\frac{B-H}{4\pi }=0,
\end{equation*}%
or

\begin{equation*}
B = H - 4\pi \frac{\partial \tilde {\varepsilon }}{\partial B} \equiv H +
4\pi M,
\end{equation*}

\noindent which gives us the expression for the magnetic moment $M(B)\equiv -%
{\partial \tilde{\varepsilon}}/\partial B$\textbf{\textit{.}} The energy $%
\tilde{\varepsilon}$ is described by the exact LK formula, which takes into
account both temperature and the Fermi surface shape but is very cumbersome.
We take the simplest approximation for $\tilde{\varepsilon}$, sufficient for
understanding the reasons of the phenomenon described, namely,%
\begin{equation*}
\tilde{\varepsilon}=a\mathrm{cos}\varphi \mathbf{,\ }
\end{equation*}%
with the phase%
\begin{equation}
\varphi =2\pi F/B\mathbf{.\ }
\end{equation}%

\noindent Here the amplitude $a$ is governed by various experimental
conditions whereas the magnetic frequency $F$ is directly proportional to
the area $A$ of extremal Fermi surface cross-section for the given metal
(see Fig. 5), as mentioned above, represents Onsager relation%
\begin{equation*}
F=c\hbar A/2\pi e.
\end{equation*}

\noindent It is easy to see that if $a\ll 1$, the difference between $B$ and 
$H$ is negligibly small as compared to the oscillation period, that is the
phase $\varphi $\textit{\ }remains practically unchanged at the replacement
of $B$ by $H.$It seems evident that in this case the first derivative of $%
\tilde{\varepsilon}$ - the magnetic moment $M$ - and its second derivative
-- the \textit{differential susceptibility}

\begin{equation*}
\chi =\mathrm{dM/dH}\cong \chi _{B}=\partial M/\partial H\equiv {\partial }%
^{2}{\tilde{\varepsilon}}/\partial B^{2}
\end{equation*}

\noindent must have the sine or cosine shape as functions of the magnetic
field. This means that the experimentally measured $M(H)$ and $\chi (H)$
dependences will have the same shape. This requirement is usually fulfilled
but under some conditions it might be definitely not the case, which is very
important for the domain formation.

From the expression for a phase (3) we have

\begin{equation*}
\Delta \varphi = \frac{\partial \varphi }{\partial H}\Delta H = - \frac{2\pi
F}{H^2}\Delta H
\end{equation*}

\noindent and for $\Delta \varphi $=2$\pi $

\begin{equation*}
\Delta H=-\frac{H^{2}}{F}.
\end{equation*}%
Here $H$ is the \textit{applied} magnetic field, and negative sign shows the
phase increase with inverse field. Note that this expression is insensitive
to the difference $B-H$ that appears and disappears periodically and always
vanishes at $\varphi $=2$\pi $n. So, one can already see that the
\textquotedblleft period\textquotedblright\ of oscillations in direct
magnetic field $\Delta H$ decreases quadratically with the magnetic field.
This means that oscillations become very \textquotedblleft
fast\textquotedblright\ at low field with a corresponding increase of the
field derivatives and the differential susceptibility $\chi _{B}$
increasing, as a matter of fact, without limit. Of course, it is possible if
and as long as the value of \textbf{$\omega \tau $} remains exceeding one
with the decrease of magnetic field, that is electrons still can perform
more than one rotation in magnetic field. Let us look at consequences of
such susceptibility growth. In the case considered, the field-induced change
in magnetic induction will be essentially different depending on the sign of 
$\chi _{B}$\textbf{\textit{.}} Indeed, this change%
\begin{equation*}
\delta B=\delta H+4\pi \delta M=\delta H+4\pi \frac{\partial M}{\partial B}%
\delta B\equiv \delta H+4\pi \chi _{B}\delta B,
\end{equation*}%
i.e.%
\begin{equation*}
1+4\pi \chi =\frac{\partial B}{\partial H}=\frac{1}{1-4\pi \chi _{B}},
\end{equation*}%
or%
\begin{equation*}
4\pi \chi =\frac{4\pi \chi _{B}}{1-4\pi \chi _{B}}.
\end{equation*}%
From this it follows that if the absolute value of $\chi _{B}$ grows, the
denominator increases, $\partial B/\partial H\rightarrow 0$ and 4$\pi \chi
\rightarrow -1$ for negative $\chi _{B}$, while for positive $\chi _{B}$ the
denominator vanishes and at $\chi _{B}\rightarrow $ 1/4$\pi $ $\partial
B/\partial H\rightarrow \infty $, so that the sample induction has to
increase like a \textit{jump}. As a result, instead of a sine-like
(harmonic) signal, the following picture should be observed in dHvA
experiments. In the vicinity of a minimum of $\tilde{\varepsilon}$, where $%
\chi _{B}$ $<0$, $B$ remains practically unchanged over almost all period,
and 4$\pi \delta M\approx -\delta M$, and 4$\pi \chi \cong $ $-1$ (almost as
in a \textquotedblleft superconductor\textquotedblright ). But in the
vicinity of a maximum of $\tilde{\varepsilon}$, where $\chi _{B}$ is
positive, the induction $B$ and, hence 4$\pi M$, 

\begin{figure}
   \centering
   \includegraphics[width=90mm,height=34mm]{fig6.gif}
   \caption{Transformation of the shape of dHvA signal $y(x)$ when
passing from $M(B)$ to $M(H)$. In the left panel, the x-axis represents the
magnetic field inside of a sample, i.e. the induction $B$. In the right
panel, the $x$-axis represents the applied magnetic field $H$, reproducing a
real magnetization measurement. In both pictures, the curves $y(x)$ are
shown (from up to down) for increasing of $a$: $a\ll 1,$ $a=1$ and $a\gg 1$,
($a$ can increase, for instance, with temperature decrease). With this
transformation, every point of the left curves is shifted by the value of
its respective $y$-coordinate either to the left ($y>0$) or to the right ($%
y<0$), according to $H=B-4\protect\pi M(B$). The most upper curve almost
does not change with this transition. Note the jumps $\Delta y=\Delta B$ on
the lower right curve.}
    \label{Fig6}
\end{figure}

increase
stepwise by $\Delta B$ approximately equal to the oscillation period (see
Fig. 6), as soon as

\begin{equation}
\chi _{B}\geq 1/4\pi .
\end{equation}

Such a \textit{saw-toothed} $M(H)$ dependence with almost vertical induction
steps was first observed by D. Shoenberg in the noble metal samples [6]. As
we see, metals in magnetic field demonstrate quite a \textquotedblleft
reasonable\textquotedblright\ behavior: the sample induction $B$ changes in
such a manner that the energy stays maximally long near its minimal value
while the high energy intervals $\Delta B$ (strictly speaking, these are the
regions of absolute instability) are jumped over (see Fig. 7).

\begin{figure}
   \centering
   \includegraphics[width=91mm,height=60mm]{fig7.gif}
   \caption{On the right: $B(H)$ dependence in a long
sample for $y(x)$ given by the lower right curve of Fig. 6. The diagonal $%
B=H $ crosses the curve at $M=0$. The function $\tilde{\protect\varepsilon}%
(B)$ is shown at the left. The regions of high energy (absolute instability)
are absent in the sample as they \textquotedblleft jump\textquotedblright\
by $\Delta B$.}
    \label{Fig7}
\end{figure}

By now, one can already \textquotedblleft guess\textquotedblright\ (all of
us are slow on the uptake) that the choice of some other geometry of
experiment, say, by using a planar sample normal to the field, rather than a
cylinder parallel to it, might provoke a different scenario of events.
Indeed, in the planar geometry with all sample dimensions considerably
exceeding its thickness, the compulsory continuity of the normal component
of $B$ results in the requirement%
\begin{equation}
B=H
\end{equation}

Since $H$ varies continuously, therefore in this geometry of experiment any
jump in $\Delta B$ cannot exist in principle. It means, in turn, that the
above-mentioned \textquotedblleft reasonable\textquotedblright\ behavior of
metal cannot presumably be realized: the induction must acquire all
consecutive values near the energy maximum, which is definitely unfavorable.
This is what to think about. Looking ahead, we declare at once that, thanks
to domains, a metal manages to behave \textquotedblleft
reasonably\textquotedblright\ and rush an unfavorable region $\Delta B$ by
in this case as well. Nevertheless, several years had passed until Condon
formulated the idea of domain formation [1].

It is to be said that this idea was preceded and, to some extent, stimulated
by numerous experiments with beryllium single crystals. The Fermi surface of
this metal contains the so-called \textquotedblleft
cigars\textquotedblright\ with the shape quite similar to a long cylinder.
That is why in this metal amplitudes of the dHvA and other effects in
magnetic field are very high. Besides dHvA, where the above-mentioned
stepwise behavior of magnetic moment is well-pronounced, many other effects,
including the transversal magnetoresistance, were measured. Very large
amplitude of these oscillations is a specific feature of the beryllium Fermi
surface. It is essential that what is measured, is a long strip or a rod
perpendicular to the external magnetic field. That is, absolutely unusual
magnetic field dependence of the amplitude of these oscillations could be
explained only by the domain formation in a sample, or, in other words,
breaking it up into areas of different magnetization.

\begin{figure}
   \centering
   \includegraphics[width=89mm,height=70mm]{fig8.gif}
   \caption{The energy variation as a
function of $B$ is shown for a small region slightly larger than one period
of oscillating function $\tilde{\protect\varepsilon}$. The external magnetic
field $H_{0}$ is chosen to be exactly at the maximum of $\tilde{\protect%
\varepsilon}$. The parabola $\mathord{\buildrel{\lower3pt\hbox{$%
\scriptscriptstyle\smile$}}\over 
{\varepsilon }}$ represents the variation
of the magnetization energy for a very long sample in applied magnetic field 
$H_{0}$. The upper curve shows the sum $\tilde{\protect\varepsilon}+ 
\mathord{\buildrel{\lower3pt\hbox{$\scriptscriptstyle\smile$}}\over
{%
\varepsilon }}$. It has minima at $B_{1}$ and $B_{2}$. The energy of a
plate-like sample with domains is shown by the dashed line.}
    \label{Fig8}
\end{figure}

To understand it better, let us appeal to a graphical presentation of the
full energy change (2) depending on the magnetic field in the sample, that
is on $B$\textbf{\textit{.}} The graph (Fig. 8) shows a small region of the
variations of $B$ in the vicinity of a given external magnetic field $H_{0}$%
, that is exactly corresponds to an immediate maximum of the oscillating
function $\tilde{\varepsilon}$. The parabola $%
\mathord{\buildrel{\lower3pt\hbox{$\scriptscriptstyle\smile$}}\over 
{\varepsilon }}=(B-H_{0})^{2}/8\pi $ depicts the second term in (2), that is
the magnetization current energy caused by the difference ($B-H_{0}$ )%
\textbf{\textit{\ }}in a sample. We emphasize that for the time being we are
dealing again with a long sample oriented along the magnetic field. The
upper curve shows $\tilde{\varepsilon}+%
\mathord{\buildrel{\lower3pt\hbox{$\scriptscriptstyle\smile$}}\over 
{\varepsilon }}$ - the total energy (2). Our figure corresponds to the
situation when the curvature of parabola is obviously less than the
curvature of $\tilde{\varepsilon}$ in a maximum, so that the condition (4)
is satisfied. Only in this case the sum $\tilde{\varepsilon}+%
\mathord{\buildrel{\lower3pt\hbox{$\scriptscriptstyle\smile$}}\over 
{\varepsilon }}$ has two symmetric minima in the points $B_{1}$ and $B_{2}$.
(In the opposite case, when 4$\pi \chi < 1$, the curve $\tilde{%
\varepsilon}+%
\mathord{\buildrel{\lower3pt\hbox{$\scriptscriptstyle\smile$}}\over 
{\varepsilon }}$ has always only one minimum). Let us remind that we have
chosen $H_{0}$ exactly in the maximum and hence the energies in these minima
coincide. Of course, if one shifts an applied magnetic field $H$ slightly
left of $H_{0}$, with a simultaneous shift of the parabola $%
\mathord{\buildrel{\lower3pt\hbox{$\scriptscriptstyle\smile$}}\over 
{\varepsilon }}$, then $\tilde{\varepsilon}+%
\mathord{\buildrel{\lower3pt\hbox{$\scriptscriptstyle\smile$}}\over 
{\varepsilon }}$ will become slightly warped, with energy in the minimum $%
B_{1}$ becoming less than in $B_{2}$. A similar right shift will cause an
opposite kind of warping and lowering the minimum $B_{2}$ below $B_{1}$.
Since the state of a metal always corresponds to minimal energy, as soon as
the external field crosses the point $H=H_{0}$, the sample magnetic
induction jumps from $B_{1}$ to $B_{2}$. The negative magnetization $%
B_{1}-H_{0}=$4$\pi M_{1}$ at this point will, respectively, change into the
positive one $B_{2}-H_{0}=$ 4$\pi M_{2}$, in other words, the sample jumps
from a dia- to a paramagnetic state.

Now we look at the domain formation in this picture (Fig. 8). To do it, we
take the same crystal with the same crystallographic orientation related to
the magnetic field, that is leave $\tilde{\varepsilon}$ unchanged, but
reshape the sample transforming it into a large thin plate perpendicular to
the field, so that, with the well known boundary condition, equality (5)
must be fulfilled everywhere. This means $B-H=0$, allowing us to remove
mentally the parabola $%
\mathord{\buildrel{\lower3pt\hbox{$\scriptscriptstyle\smile$}}\over 
{\varepsilon }}$. As a result, the sum $\tilde{\varepsilon}+%
\mathord{\buildrel{\lower3pt\hbox{$\scriptscriptstyle\smile$}}\over 
{\varepsilon }}$ (2) simply coincides now with $\tilde{\varepsilon}$. By
comparing this result with the previous one, that is with the curve $\tilde{%
\varepsilon}+%
\mathord{\buildrel{\lower3pt\hbox{$\scriptscriptstyle\smile$}}\over 
{\varepsilon }}$ on the figure, we see that over a large range of magnetic
fields in the vicinity of $H_{0}$, the energy of metal becomes considerably
higher than the minimal value realized in a thin sample. This exceeding is
maximal at $H=H_{0}$ being equal to \textit{$\delta \varepsilon $}. The
question is if it is possible to reduce the energy by dividing our plate
into a set of \textit{thin} regions -- \textquotedblleft
domains\textquotedblright . Let their length, which is the plate thickness,
be much larger than the \textquotedblleft domain\textquotedblright\
thickness, in which case the cross-section of such a \textquotedblleft
domain\textquotedblright\ looks similar to that for a solitary long sample
oriented along the field. That is why we can apply the formula (2) that is
the curve $\tilde{\varepsilon}+%
\mathord{\buildrel{\lower3pt\hbox{$\scriptscriptstyle\smile$}}\over 
{\varepsilon }}$ for each of them. If we now break them up into two sorts
with the induction values $B_{1}$ and $B_{2}$ and mix carefully to make
these sorts alternating everywhere, then these domains (already without
quotation marks) will represent what is called Condon's domains. Since
domain size and number are for both sorts equal, in each sample region much
larger than the domain size the average induction remains equal to $H_{0}$,
that is the condition (5) is now satisfied on average throughout the whole
sample. In each domain the energy (2) corresponds to a minimum, which means
that for the whole plate the energy gain will be the same, namely \textit{$%
\delta \varepsilon $}. Fig. 9 schematically shows such domain structure. If
the magnetic field changes towards $B_{1}$ or $B_{2}$, then the domain sizes
vary correspondingly, increasing the thickness of one sort of domains and
decreasing the other in such a way that the condition (5) remain on average
satisfied. Simple calculations show that formation of domains with the
constant values $B_{1}$ and $B_{2}$ in each sort of them becomes more
profitable than the uniform state for all values of magnetic field in region 
$B_{1}<H<B_{2}$. This energy gain is shown in the Fig. 8 by the dashed line.%

\begin{figure}
   \centering
   \includegraphics[width=82mm,height=31mm]{fig9.gif}
   \caption{Schematic picture of the domain structure in a plate-like sample. In reality, the period of the structure is much smaller than the thickness $d$. The arrows show directions of magnetization in phases 1 and 2.}
    \label{Fig9}
\end{figure}

In 1966 Condon formulated first the idea of such domains, and already in two
years Condon and Walstedt [7] demonstrated the domain formation in nuclear
magnetic resonance (NMR) experiments in silver. Let us remind that nuclear
magnetic moment, or \textit{spin}, rotates in the magnetic field with the
angular frequency of such rotation, or \textit{precession}, strictly
proportional to the field strength. If, besides, an additional
high-frequency electromagnetic field effects a nucleus, a \textit{resonant}
absorption of electromagnetic energy by this nucleus occurs as soon as the
field frequency coincides with the precession frequency. The frequency of
a.c. field, created usually by a small coil winded sometimes directly on a
sample, can be measured with enormous precision, and hence NMR gives the
opportunity to measure magnetic field in a medium with the same precision
(of course, only if one succeeds in measuring the absorption itself, which
is not a simple task). In a uniform field, a narrow NMR line is observed,
whereas any non-uniformity broadens this NMR peak (line). Condon and
Walstedt were the first who observed \textit{two} coexisting resonant
frequencies, that is the line \textit{splitting.} At changing the external
magnetic field the effect arose periodically with the period corresponding
to the dHvA period in the same sample, while the magnitude of splitting had
the order of half a period and corresponded to two sorts of domains with
induction values $B_{1}$ and $B_{2}.$

Unqualified success of this experiment was a well-deserved result of
overcoming a large number of difficulties. Besides all already mentioned
conditions of domain formation including low temperature of 1.4 K, very high
magnetic field uniformity with spatial fluctuations essentially less than
the splitting value $\delta B=B_{2}-B_{1}=12$ Gauss against the field of 9
T, and very high perfection of a Ag single crystal, an additional difficulty
consisted in detecting of NMR in a metal, especially in a very pure metal,
as in the experiment. The case is that a.c. electromagnetic field penetrates
a metal only at very small depth, the so-called \textit{skin} layer. That is
why only small number of nuclei near the sample surface contribute to
absorption. With the account of all above-mentioned factors, the authors
presumably had very few chances for success, thus the result speaks for
itself.

The authors naturally tried to obtain the same result for beryllium, the
\textquotedblleft champion\textquotedblright\ among metals with the highest
amplitude of the dHvA effect, but suffered a reverse. The method did not
work. Contrary to silver where the nuclear magnetic moment is equal to $1/2$
and its projection to the magnetic field has only two allowed values: along
and opposite to the field, this moment for beryllium is equal to $3/2$, so
that initially, without any domains, the so-called \textit{quadrupole }%
splitting of the NMR line already exists. This is one more difficulty in
observation of domains in many metals by the NMR method. All the problems
discussed where presumably the reason why after a success with silver and
failure with beryllium no single work devoted to revealing diamagnetic
domains by the NMR method appeared in the literature. The progress and all
recent achievements in visualization of diamagnetic domains is related to a
new investigation method -- Muon Spin Rotation, called $\mu $SR [8]

This method was developed at the \textquotedblleft
interface\textquotedblright\ between two branches of physics -- nuclear
physics and condensed matter physics, and actually is almost complete analog
of the NMR. As early as in 1979, Yu. Belousov and V. Smilga suggested to use
it for observation of Condon domains [9]. The technique of that time,
however, was not yet adequate, \textquotedblleft
interface\textquotedblright\ was not formed, and their work remained, alas,
unnoticeable. 16 years later the idea of using the $\mu $SR method for
domain observation was reborn, this time with a project proposed by G. Solt
at the Paul Scherrer Institute, Switzerland, where this method is of a wide
use. Experiments in beryllium were a success [10], and splitting of the $\mu 
$SR peak, similar to that for NMR, caused by diamagnetic domain formation,
was observed.

The $\mu $SR method, in spite of its direct analogy with NMR, has, of
course, a number of distinctions as well. Muons are unstable elementary
particles with the lifetime close to two microseconds. They represent an
outcome of activity of a powerful accelerator. A positive muon, having
sufficiently high initial energy, can penetrate the sample at fairly large
depth and stop at some interstitial remaining there during the whole
lifetime. It also has a spin precessing in exact correspondence with the
local value of magnetic field. Decay of a muon creates a positron, or
anti-electron, which rushes our mostly in the direction of its spin and is
registered by one or another detector. In the experiment, a great number of
muons is detected, with all their spins rotating from strictly the same
starting position. If all muons are in the same magnetic field, then the
number of registered events in each detector will oscillate with time with
the precession frequency $f$ exactly determining the magnitude of this
magnetic field, that is $f=gB$, where the constant $g$ is well known for
muon. In this methods there is no need in a.c. electromagnetic field since
the precession frequency is measured directly, and therefore the first
difficulty of NMR measurements caused by a skin-layer no longer exists. The
second difficulty is absent as well since in any matrix the
\textquotedblleft job\textquotedblright\ was made by the same
\textquotedblleft test\textquotedblright\ instrument with the spin $1/2$.
The fact that spin precession occurs far enough from the sample surface,
represents the third important advantage of this method, at least for this
problem. As a result, by analogy with NMR, the width of $\mu $SR peak
corresponds to the amplitude of magnetic field non-uniformity. If now the
sample becomes stratified into two phases with the magnetic field values $%
B_{1}$ and $B_{2}$, that is domains, then one part of muons will find
themselves in the field $B_{1}$ and the other part -- in the field $B_{2}$,
which will result in two precession frequencies and, respectively, in a
splitting of the $\mu $SR peak into two peaks.

\begin{figure}
   \centering
   \includegraphics[width=87mm,height=56mm]{fig10.gif}
   \caption{Several $%
{\mu}%
$SR spectra over a small region of applied magnetic field $H$ near the onset
of Condon domains. (The spectra are shown without noise). The peaks of the
spectra without domains are always along the diagonal $B=H$. In the domain
region, there are two peaks of $f_{1}$ and $f_{2}$ corresponding to $B_{1}$
and $B_{2}$.}
    \label{Fig10}
\end{figure}

The Fig. 10 demonstrates the results of $\mu $SR experiment on a crystalline
plate of beryllium [11]. Each time when $H$ goes through the region $%
B_{1}<H<B_{2}$, the spectrum will split into two peaks with the fixed
frequencies corresponding to $B_{1}$\textbf{\textit{\ }}and $B_{2}$\textbf{%
\textit{.}} While the field changes, the amplitude of one peak decreases and
the amplitude of the other increases, which corresponds to the change of
relative volumes occupied by these two phases. Analysis of the data
available confirms that the relationship (5) is always exactly fulfilled. At
any other values of magnetic field beyond the given range, a standard narrow
peak is observed with the frequency corresponding to this field.

Now we can say that a successful result of the experiments with beryllium is
quite natural since this material, as it has been already mentioned, is a
\textquotedblleft champion\textquotedblright\ in the dHvA amplitude. In the
magnetic field 3T, diamagnetic domains exist up to the temperature $\sim $ 3K. 
For the most other metals, however, the dHvA amplitude is
considerably less than for beryllium. As a rule, it remains under all
conditions noticeably less than one tenth of period. But it is well known,
that the condition (4) or $a\geq $1 (which is the same for a sine-shape dHvA
signal) is satisfied if the amplitude $4\pi M\geq P/2\pi $ ($P$ is the
period). At first glance, Condon domains seem impossible under these
conditions representing a very rare phenomenon, which is, however, not the
case. Actually, the shape of dHvA signal at very low temperatures is already
essentially different from a sine curve. A careful analysis shows that even
for a very small dHvA amplitude, the condition (4) at sufficiently low
temperatures will be satisfied without fail, though the range $B_{2}-B_{1}$
in this case may be very narrow, essentially less than the period.
Therefore, the difference in magnetization and the $\mu $SR peak splitting
are extremely small, and experimental observation of domain formation
becomes much more difficult. It requires both absolutely perfect crystals
and more sophisticated measuring technique.

Just such experiments have been recently performed in the same the Paul
Scherrer Institute. The mentioned more sophisticated technique making it
possible to diminish essentially the noise level, the so-called \textit{MORE}%
, was worked up at this institute. Formation of diamagnetic domains was
discovered in all measured single crystals of tin, aluminum, indium and
lead. (They were grown in the P.L. Kapitsa Institute of Physical Problems
almost 30 years ago). The condition for domains to exist was restricted to
several tenth of Kelvin temperature. Success of this work [11] was naturally
based on the many year work of numbers and numbers of physicists. Now one
can be sure that diamagnetic, or Condon, domains represent a phenomenon
spread as widely as the dHvA effect though requiring much more rigid
conditions for their observation.

Two more questions should be mentioned, at least casually, in this paper.
The first, quite natural question is that of a mechanism of electric current
in a rather thin, of order of one micron, domain wall. So, in beryllium at $%
B_{2}-B_{1}$ around 30 G, the current density in the wall has to be $\cong
3.10^{5}$\`{A}/cm$^{2}$. It is a very large value. In ordinary magnetic
domain formed by spins (atomic magnetic moments) with opposite direction,
this mechanism is clear: currents circulating in adjacent domains,
respectively, clockwise and anticlockwise, add at the boundary forming thus
a magnetization current. But in our case Larmor rotation of electrons is
identical both sides of the boundary so that in this sense the boundary is
not marked out. The answer consists in the fact that in the dHvA effect not
only magnetization but also the crystal size is varied, which is knows as
the \textit{striction }effect. The case is that the phase of oscillating
energy $\varphi $ =2$\pi F/B$ (3) is determined not only by the induction $B$
but also by the value of $F$, that is by the Fermi surface cross-section,
which, in turn, depends on the sizes of a crystal cell. That is why a metal
in the external magnetic field changes not only its magnetization but also
the cross-section, or the volume, of the Fermi surface by a proper dimension
changes, in order to approach the energy minimum \textquotedblleft
faster\textquotedblright . While jumping from $B_{1}$ to $B_{2}$, this
striction also suffers a jump. In this process, opposite magnetization
corresponds, in some sense, to opposite deformation. In domains, on the
contrary, variation of deformation from one value to the other must occur in
a domain wall more or less smoothly. Now it is clear that larger Fermi
surface volume, that is larger charge density, corresponds to a diamagnetic
phase while smaller volume corresponds to a paramagnetic phase with the
charge density gradient in the domain wall providing the required
magnetization current. Of course, striction is directly proportional to
magnetization and has a very small magnitude. So, in the above-mentioned
beryllium with a record magnitude of effects the deformation has the order
of one per million. Thus, formation of the domain structure is also
accompanied by a corresponding, unfortunately very small, periodic
deformation of the unite cell size and, moreover, relief at the sample
surface. This makes it very difficult for observation even by a heavily
aided eye.

This is not the only place where deformation, or formation of domains from
different density phases, reveals itself. Measurements of magnetostriction
in a beryllium plate resulted in discovery of an absolutely amazing property
of the formed domain structures, which cannot be named other than
``supersoftness'' [12]. It should be noted that beryllium by itself is a
very hard metal, inferior by this property only to tungsten and iridium. Its
Young's modulus is almost one order of magnitude higher than that of copper.
Nevertheless, a copper needle of a regulating screw pressing the beryllium
plate to the measuring instrument with minimal force, periodically, at the
formation of domain structure, \textit{comes down} the sample at a rather
noticeable depth. The depth of a ``pit'' under the needle, which, of course,
heals instantly as soon as the sample becomes single-phased, corresponds to
at least \textit{hundredfold} drop in the Young's modulus. This unique
behavior can be explained only in terms of corresponding domain
restructuring in the vicinity of the needle.

The second question is as follows. In our opinion, there is a direct analogy
between described diamagnetic domains and alternation of normal and
superconducting phases in the known \textit{intermediate }state of a type I
superconductor. Indeed, a long thin cylindrical superconducting sample
oriented along the magnetic field, at some critical field $H_{c}$
demonstrates a \textit{stepwise} transition from the superconducting state
with $B_{1}=0$ into the normal one with $B_{2}=H_{c}$. Of course, both
states correspond to a minimum of energy. If we now take a sample of the
same metal but in the shape of a thin plate perpendicular to the field, pure
geometric considerations will again bring us to the necessary condition (5).
However, in the magnetic field interval between $B_{1}$ and $B_{2}$, that is
between zero and $H_{c}$, a uniform solution will possess some excess
energy. Minimal energy will be achieved by fragmentation of a sample into
alternating \textquotedblleft domains\textquotedblright\ with the induction $%
B_{1}$=0 and $B_{2}=H_{c}$, that is into normal and superconducting phases,
exactly as in the case of diamagnetic domains. In this case the condition
(5) is again fulfilled on average on the account of proportional changes of
the relative volumes of the phases. Actually the analogy is even closer.
From the analysis of domain structure periods it results that these may be
very close in samples of the same thickness. This means that domain
structures of either shape may be rather similar to such different phenomena
as superconductivity and dHvA effect. Unfortunately, this is the end of
analogy and remaining distinctions have a fundamental character. If the
\textquotedblleft magnetic contrast\textquotedblright , that is the ratio of 
$B_{2}-B_{1}$ to $B_{2}$, is almost hundred percent for the intermediate
state image, then for Condon domains it is so far 0.1{\%} at best. Besides,
the magnetic field itself is here hundred times more, which creates an
additional obstacle for the magnetooptical method used for imaging. However,
the principal possibility of obtaining a diamagnetic domain image remains,
which gives ground for some optimism.

In conclusion, a couple of words should be said regarding \textquotedblleft
practical application\textquotedblright\ of Condon domains. They give an
absolutely unexpected possibility of direct approach to the question of
compressibility of metals. It appears that if compressibility of metals $%
\beta _{\mathrm{met}}$ is governed exclusively by the kinetic energy of the
electron gas, i.e. $\beta _{\mathrm{met}}=\beta _{\mathrm{el},}$ then only
in this case no contact voltage exists between domains and, hence, the
domain wall interior contains no electric field.

In 1957 M.I. Kaganov, I.M. Lifshits, and K.D. Sinel'nikov predicted
theoretically [13] the effect of Fermi level oscillations with magnetic field

\begin{equation*}
\delta \mu _{KLS} (H) = \frac{\partial \tilde {\varepsilon }(H)}{\partial N},
\end{equation*}

\noindent where the energy $\tilde{\varepsilon}$ already mentioned above, is
described by the exact LK formula and $N$ -- is the density of electrons.
The result was obtained for the case of constant $N$. Nevertheless, as a
result of striction, the volume $V$ changes, $N=N_{0}/V,N_{0}$ is a constant
quantity of electrons in a crystal, and the complete change of Fermi level is

\begin{equation*}
d\mu = \delta \mu _{KLS} + \frac{\partial \mu }{\partial V}\delta V,
\end{equation*}

\noindent where $\delta V$ is the striction in the crystal. We can write%
\begin{equation*}
\delta \mu _{KLS}=\frac{\partial \tilde{\varepsilon}}{\partial V}\frac{%
\partial V}{\partial N}=\frac{V}{N}\delta \tilde{p},
\end{equation*}%
where%
\begin{equation*}
\delta \tilde{p}=-\frac{\partial \tilde{\varepsilon}}{\partial V},
\end{equation*}%
and

\begin{equation*}
\delta V = \frac{\partial V}{\partial p}\delta p\ast ,
\end{equation*}

\noindent where $\delta p$* is the pressure decrease caused by striction.
The total variation of pressure for a sample with free boundaries is zero%
\begin{equation*}
\delta \tilde{p}+\delta p\ast =0,
\end{equation*}%
and

\begin{equation*}
d\mu =\frac{V}{N}\delta \tilde{p}(1+N\frac{\partial \mu }{\partial V}\beta _{%
\mathrm{met}}).
\end{equation*}%
Here the $\beta _{\mathrm{met}}$ is a compressibility of the metal which can
be found in a Handbook. Now we rewrite%
\begin{equation*}
\frac{\partial \mu }{\partial V}=-\frac{\partial p}{\partial N},
\end{equation*}%
and from%
\begin{equation*}
\frac{\partial N}{\partial p}=\frac{\partial (N_{0}/V)}{\partial p}=-\frac{%
N_{0}}{V^{2}}\frac{\partial V}{\partial p}=-\frac{N}{V}\frac{\partial V}{%
\partial p}=N\beta _{\mathrm{el}}
\end{equation*}%
we have

\begin{equation*}
\frac{\partial \mu }{\partial V}=-\frac{1}{N\beta _{\mathrm{el}}}.
\end{equation*}%
Here $\beta _{\mathrm{el}}$ is a compressibility of electron gas which could
be found, in principle, from an equation of state for electron gas in this
metal. Of course, $\beta _{\mathrm{el}}$ is connected with a kinetic energy
of electrons only. At least, we have for the net shift of Fermi level

\begin{equation*}
d\mu =\beta \mu _{KLS}(1-\frac{\beta _{\mathrm{met}}}{\beta _{\mathrm{el}}}).
\end{equation*}

Therefore, one can, in principle, find the value of $\beta _{\mathrm{el}}$
from a contact voltage measurement. In the case $\beta _{\mathrm{met}}=\beta
_{\mathrm{el}}$ the first derivative of exchange energy has a maximum and
only kinetic energy of electrons contributes in compressibility. Only in
such case the effect of Fermi level oscillations is wholly compensated as a
result of magnetostriction gaining no contact voltage between domains, no
electrical field in domain walls, and no extra energy. Maybe just for this
not trivial reason we can see the Condon domains in all metals [11].

On the other hand, if we assume that the whole magnetization current in a
domain wall is a result of only charge density gradient, that is domains are
actually only diamagnetic, with negligible role of spins, it appears that
compressibility of metals is completely determined by the construction of
their Fermi surface.

Indeed, let the difference in magnetization between neighboring domains be
really caused by deformation accompanied by electron density changes. Then
the magnetization current density in a domain wall can be described by the
formula [14]

\begin{center}
$j_{m}=c$ curl $\sum\limits_{k}{n_{k}(r)\mu _{k}}$
\end{center}

Here, $n_{k}(r)$ is the number of Larmor orbits corresponding to the $\mu $
magnetic moment of a unit volume $\mu _{k}$, $c$ is the light velocity. Let
us integrate j$_{m}$ over the domain wall thickness from one domain to
another taking into account that the orbital magnetic moments of all
electrons are parallel to the magnetic field. This gives the magnetization
current in the wall related per unit length of this wall along the field,

\begin{center}
$J$= $c\sum\limits_{k}{(N_{2}-N_{1})_{k}\mu _{k}}$
\end{center}

\noindent where $(N_{1},_{2})_{k}$ are the volume densities of charges with
magnetic moment $\mu _{k}$ in neighboring domains. Since the $\delta N$
difference is small, all orbits can be considered to be situated on the
Fermi surface. The characteristic values can be estimated as follows. The
magnetic moment of a Larmor orbit is

\begin{equation*}
\mu _k = \frac{J_L S_L }{c},
\end{equation*}

\noindent where $J_{L}=\omega _{c}e/2\pi $ is the current on Larmor orbit
and $S_{L}=\pi R_{H}^{2}$ is its area. Here $\omega _{c}=eH/mc$ is the
cyclotron frequency, $e$ is the charge of the electron, $R_{H}=v_{\bot }$/$%
\omega $ is the Larmor radius, and $v_{\bot }$ is the velocity of electrons
on the Fermi surface in the direction normal to the field. We can write the
complete current $J$ in the domain wall per unit wall length in the magnetic
field direction as

\begin{equation*}
J = \delta N\frac{c}{H}\frac{mv_F^2 }{2}C_{FS} = \delta N\frac{c}{H}%
\varepsilon _F C_{FS} ,
\end{equation*}

\noindent where $\delta N$ is the total difference of the numbers of charge
carriers (electrons and holes) in neighboring domains, that is, the
difference of the Fermi surface volumes in these domains, and the constant $%
C_{FS}$ is a result of averaging $v_{\bot }$ over the Fermi surface. As the
induction jump between neighboring domains is%
\begin{equation*}
\Delta B=4\pi \Delta M=(4\pi /c)J,
\end{equation*}

\noindent where $J$ is just the current in the domain wall, we have%
\begin{equation*}
\Delta MH=\delta N\varepsilon _{F}C_{FS}.
\end{equation*}%
The changing of charge density $\delta N$/$N_{0}$ can always be considered
equal to $\gamma C_{\gamma }$ where $\gamma $ is striction and $C_{\gamma }$
is the coefficient unambiguously determined by the Fermi surface shape.
(Clearly, this coefficient is equal to 3 in the model of free electrons).
So, we can rewrite%
\begin{equation*}
\Delta MH=\gamma N_{0}\varepsilon _{F}C_{FS}C_{\gamma }.
\end{equation*}%
At least, we have the well-known formula for striction [6]

\begin{equation*}
\gamma =\frac{MH}{Y}\frac{\partial \ln A}{\partial \gamma }=\frac{MH}{Y}%
C_{A}.
\end{equation*}%
Here $Y$ is Young's modulus and the constant $C_{A}$ shows a changing of
Fermi surface cross-section $A$ with striction $\gamma $. As a result, we
have%
\begin{equation*}
Y=N_{0}\varepsilon _{F}C_{FS}C_{\gamma }C_{A},
\end{equation*}

\noindent where all coefficients are fully determined by the Fermi surface
structure. Here, $\varepsilon _{F}$ is the kinetic energy of electrons on
the Fermi surface, that is, $\varepsilon _{F}=\hbar ^2k_F^2 / 2m$. For
instance, for beryllium [12], the correct Young's modulus value was obtained
in such a simple way.

To summarize, we can say that no wonderful Condon domains are connected with
a compressibility of metal for its appearance is directly connected with
deformation. But the concept described above in a very simple way, shows
that conduction electrons should \textit{fully} determine its
compressibility coefficient. Of course, it is much more strong result and
the result is strange. At the same time, the formation of diamagnetic
domains is doubtless characteristic for all metals; the only problem is the
extremal difficulty of creating the necessary conditions for most of them.
As mentioned, such domains were observed in silver, beryllium, tin, lead,
indium and aluminum. In other words, the very possibility of the existence
of diamagnetic domains lends support to the point of view according to which
conduction electrons should fully, or almost fully determine the
compressibility of metals. Of course, it is very difficult to say now to
what extent this conclusion is quantitatively accurate.

\bigskip

I am indebted to L. Maksimov, D. Sholt, V. Mineev, A. Dyugaev for
interesting discussions of the questions touched upon and to M. Schlenker
for useful remarks.

\newpage

\end{document}